\documentstyle[12pt]{article}

\if@twoside
\else
\fi
\begin{document}
\renewcommand{\theequation}{\thesection.\arabic{equation}}

\title{
Supersymmetric Quantization of Gauge Theories}

\author{ \sc{F G  Scholtz$^a$ and S V Shabanov$^b$\thanks
{On leave from: Laboratory for
Theoretical Physics,
JINR, Dubna, Russia.}}\\
\sc{$^a$  \small{\it Department of Physics, University
of Stellenbosch,}}\\
\sc{\small{\it 7600 Stellenbosch, South Africa}}\\
\sc{$^b$ \small{\it Department of Theoretical Physics, University of
Valencia,}}\\
\sc{\small{\it Valencia, Spain}}}

\date{8 August 1996 }

\maketitle

\begin{abstract}
We develop a new operator quantization scheme for gauge
theories in which the dynamics of the ghost sector is described by a $N=2$
supersymmetry.  In this scheme no gauge condition is imposed on the gauge
fields.  The corresponding path integral is explicitly Lorentz invariant and,
in contrast to the BRST-BFV path integral in the Lorentz gauge, it is free of
the Gribov ambiguity, i.e., it is also valid in the non-perturbative domain.
The formalism can therefore be used to
study the non-perturbative properties of
gauge theories in the infra-red region (gluon confinement).
\end{abstract}

\vspace{0.5cm}

\section{ Introduction.}
\setcounter{equation}0

Systems with first class constraints,
and in particular
relativistic gauge theories, posses
unphysical degrees of freedom which have to
be eliminated in the quantization process {\cite {dir}}.  It is, however,
important to note that in general
the process of gauge fixing and quantization does
not commute as was exemplified in {\cite {ncom}}, and that the
unphysical degrees of freedom have to be eliminated after quantization.
The conventional procedure of the path integral quantization
of gauge theories {\cite{fad}} relies on the elimination of unphysical
(gauge) degrees of freedom on the classical level via gauge fixing
and, therefore, should be modified {\cite{sha1,sha2}}.

In relativistic gauge theories the situation is
slightly more complicated since
one wants to build a Lorentz invariant theory.  To this end one has to use
Lorentz invariant gauges, e.g., the Lorentz gauge.
This implies, however, that
unphysical bosonic degrees of freedom are
dynamically activated and have to be
suppressed.  A general procedure for doing this, based on the BRST symmetry,
was given in {\cite {bv}}.

A problem common to all quantization procedures based on imposing a gauge
condition on the gauge fields, is the Gribov ambiguity {\cite {grib}}.  This
states that it is impossible (assuming certain boundary conditions at spatial
infinity) to find a gauge condition that would eliminate all gauge equivalent
configurations {\cite{sing,sol}}.  This implies that even after gauge fixing
there is still a set of residual gauge transformations under which the action
will be invariant.  Although these residual transformations do not lead to a
further reduction in the number of physical degrees of freedom, they are
important in determining the topological structure of physical configuration
(phase) space and therefore they play a crucial role in the quantization
process and the corresponding functional representation {\cite {sha2}}.  For
normal Lorentz invariant gauges, such as the Lorentz gauge, this set of
residual transformations is extremely complex {\cite{sol}} and the task of
incorporating them in the quantization process seems hopeless.

There is good reason
to believe that the Gribov ambiguity is irrelevant in the
perturbative (asymptotic free) domain {\cite {grib}}.  However, there is
evidence that it plays an important role in the infra-red, non-perturbative
behavior of the gluon propogator {\cite {grib,zwan,ber}}.

From the previous introductory remarks it should be clear that one needs to
develop a quantization procedure which eliminates the unphysical degrees of
freedom while it (a) avoids imposing a gauge condition on the gauge fields,
i.e., the Gribov ambiguity and (b) leads to a Lorentz covariant theory.  The
basic ingredients for such a scheme were given in {\cite {sch}}.  The idea of
{\cite{sch}} is to extend the theory in a supersymmetric and gauge invariant
way by introducing bosonic and fermionic ghosts transforming in a gauge
multiplet.  The supersymmetry insures that the contributions of the boson and
fermion ghosts cancel in the partition function, while the gauge invariance
makes it possible to avoid imposing
a gauge condition on the gauge fields, but
rather to eliminate the unphysical degrees of freedom by imposing a gauge
condition on the bosonic ghost fields.  Finally the whole procedure respects
the Lorentz invariance of the theory.  This program was, however, performed
within the functional integral setting and,
given the ambiguities that arise in
the functional integral formalism, it is highly desirable to develop this
program on the operator level and
derive the corresponding functional integral
representation from there.

Our aim with the present paper is to develop
this operator quantization scheme.
The steps we follow in doing this are essentially the same as outlined above.
We show how any quantum mechanical system (regardless whether it has
a gauge symmetry or not) can be extended by adding $N=2$
supersymmetric ghosts.
In the extended theory physical states are
identified as those invariant under
SUSY transformations.  Matrix elements of any system operator
calculated in the
physical subspace coincide with those of the original system.

In the case of gauge theories this extension is done by putting the ghosts in a
gauge multiplet.  This implies (a) that we not only modify the
Hamiltonian, but
also the constraints and (b) that the extension respects the gauge symmetry,
i.e, commutators of the extended Hamiltonian with the extended constraints and
the extended constraints with each other vanish weakly.
The presence of scalar
ghosts is then exploited to
remove the gauge arbitrariness by imposing gauge fixing on them.

We organize the paper as follows: In section 2 we consider a 1-dimensional
quantum system to illustrate the supersymmetric extension and to construct the
functional integral representation of the system transition amplitude in the
extended space.  In section 3 a simple mechanical model with a gauge symmetry
is considered to show how the ghosts can be added to the theory and how gauge
fixing of the variables describing the original gauge system can be avoided.
We emphasize that the choice of simple mechanical models to illustrate the
procedure is only for the convenience of presentation since the generalization
is straightforward.
In section 4 the scheme is applied to Yang-Mills theories.
The implementation of the scheme to evaluate Green's functions
is given in section 5.
Section 6 discusses the relation to normal gauge fixing and
section 7 contains
our conclusions.

\section{ Ghost extension of a quantum system.}
\setcounter{equation}0

Consider a 1-dimensional quantum
system with Hamiltonian
\begin{equation}
\hat H_s=\frac{\hat p^2}{2}+V(\hat x)\;,\quad [\hat x,\hat p]=i\,.
\label{sh}
\end{equation}
We denote by $|s\rangle$ (or $|\psi\rangle_s$) vectors in the system
Hilbert space.

Consider the Hamiltonian
\begin{equation}
\hat H_{gh}=\hat p_\eta^\dagger \hat p_\eta+\hat p_z^\dagger\hat p_z
+\omega^2(x)
(\hat z^\dagger \hat z+\hat \eta^\dagger \hat \eta)\ .
\label{2}
\end{equation}
Here $(\hat\eta^\dagger)^2=\hat \eta^2=(\hat p_\eta^\dagger)^2=\hat
p_\eta^2=0$, i.e., they are Grassmann canonical operators, while $\hat z$,
$\hat p_z$ and their adjoints have boson statistics:
\begin{equation}
[\hat z,\hat p_z^\dagger]=[\hat z^\dagger,\hat p_z]=i\quad,\quad
[\hat \eta,\hat p_\eta^\dagger]=-[\hat \eta^\dagger,\hat p_\eta]=i\,.
\label{3}
\end{equation}
Here $[\;,\;]$ is the supercommutator
$[A,B]=AB-(-1)^{\epsilon_A\epsilon_B}BA$ where $\epsilon_{A,B}$ are
Grassmann parities of $A$ and $B$, i.e., $\epsilon=0$ for bosons and even
elements of the Grassmann algebra and $\epsilon=1$ otherwise.
We shall call these variables ghosts.

We denote by $|gh\rangle^x$ (or $|\psi\rangle_{gh}^x$) vectors in the
(indefinite) Hilbert space of system (\ref{2}).  Note that $\hat H_{gh}$
depends on a real parameter $x$, therefore the index $x$ on the
states.

We construct a coordinate (Schr\"odinger) representation of the algebra
(\ref{3}) on the space of functions
$\psi_{gh}(\kappa)=\langle\kappa|\psi\rangle_{gh}^x=\langle
z,z^*,\eta,\eta^*|\psi\rangle_{gh}^x$ with the inner product
\begin{equation}
{}^x_{gh}\langle\phi|\psi\rangle_{gh}^x=\int
d\kappa\,\phi_{gh}^*(\kappa)\psi_{gh}(\kappa)\,, \label{4}
\end{equation}
where $d\kappa=dz^*dzd\eta^*d\eta$.  It is easily verified that
\begin{eqnarray}
\hat\eta&=&\eta\;,\quad\hat\eta^\dagger=\eta^*\;,\quad\hat
p_\eta=-i\frac{\overrightarrow\partial}{\partial\eta^*}\;,\quad
p_\eta^\dagger=i\frac{\overrightarrow\partial}{\partial\eta}\;,\label{cr1}\\
\hat z&=&z\;,\quad\hat z^\dagger=z^*\;,\quad\hat
p_z=-i\frac{\partial}{\partial z^*}\;,\quad
p_z^\dagger=-i\frac{\partial}{\partial z}
\label{cr2}
\end{eqnarray}
in this representation.  Note that the dagger denotes the adjoint with
respect to the inner product (\ref{4}) so that the Hamiltonian (\ref{2}) is
self-adjoint.  The inner product (\ref{4}) is indefinite and the space contains
zero and negative norm states.  We remark that negative norm states always
occur when quantizing
Grassmann Lagrangians bilinear in generalized velocities, which is the case
in our fermionic ghost sector
(see (\ref{2.28})) as well as in the conventional BRST-BFV formalism.

The ghost Hilbert space can also be constructed as a Fock space. Define
annihilation operators
\begin{eqnarray}
\hat b_1=(\omega\hat z+i\hat p_z)/\sqrt{2\omega}\;,&\quad&
\hat b_2=(\omega\hat z^\dagger+i\hat p_z^\dagger)/\sqrt{2\omega}\;,\\
\hat c_1=(\omega\hat \eta+i\hat p_\eta)/\sqrt{2\omega}\;,&\quad&
\hat c_2=(\omega\hat \eta^\dagger+i\hat p_\eta^\dagger)/\sqrt{2\omega}\;,
\label{7}
\end{eqnarray}
and creation operators through their adjoints.  From (\ref{3}) follows
\begin{equation}
[\hat b_1,\hat b_1^\dagger]=[\hat
b_2,\hat b_2^\dagger]=[\hat c_1,\hat c_1^\dagger]=1\;,\quad
[\hat c_2,\hat c_2^\dagger]=-1\,. \label{8}
\end{equation}

Defining the vacuum, $|0\rangle^x_{gh}$,
as the state annihilated by all annihilation
operators and with $ ^x_{gh}\langle 0|0\rangle^x_{gh}=1$,
one notes from (\ref{8}) that states
containing $\hat c_2^\dagger$ have negative norm.  In Fock space the ghost
Hamiltonian (\ref{2}) is given by
\begin{equation}
\hat H_{gh}=\omega\hat N_{gh}=\omega(\hat b_1^\dagger \hat b_1+
\hat b_2^\dagger \hat b_2+\hat c_1^\dagger \hat c_1-\hat c_2^\dagger
\hat c_2)\,,
\label{9}
\end{equation}
where $\hat N_{gh}$ is the ghost number operator;
it is non-negative, $\hat N_{gh}\geq 0$.

Define the fermion operators
\begin{equation}
\hat Q=\hat c_1^\dagger \hat b_1-\hat b_2^\dagger\hat c_2\,,\quad
\hat R=\hat c_1^\dagger \hat b_1+\hat b_2^\dagger\hat c_2
\label{ssc}
\end{equation}
and their adjoints $\hat Q^\dagger$ and $\hat R^\dagger$. We have
\begin{equation}
\hat N_{gh}=[\hat Q,\hat R^\dagger]=[\hat Q^\dagger,\hat R]
\label{10}
\end{equation}
and
\begin{equation}
[\hat N_{gh},\hat Q]=[\hat N_{gh},\hat Q^\dagger]=[\hat N_{gh},\hat R]=
[\hat N_{gh},\hat R^\dagger]=0\,.
\end{equation}
The ghost Hamiltonian exhibits an $N=2$ supersymmetry generated by $\hat Q$,
$\hat R$ and their adjoints.

Introduce a system with the extended Hamiltonian
\begin{equation}
\hat H=\hat H_s+\hat H_{gh}\;,\label{11}
\end{equation}
where $\hat H_{gh}$ was defined in (\ref{2}) and $\omega^2=\omega^2(\hat x)$.
Therefore states in the ghost Fock space depend on the operator $\hat x$ and
accordingly act as operators (functions of $\hat x$) on the system Hilbert
space.  This motivates the following definition of the extended space:
\begin{equation}
|\psi\rangle\rangle=|gh\rangle^x\cdot|s\rangle\,,
\end{equation}
where the dot stands to emphasize that $|gh\rangle^x$ is an operator
acting on the system state $|s\rangle$.

In the coordinate representation $|\kappa\rangle\cdot |x\rangle$,
$\hat\kappa|\kappa\rangle=\kappa|\kappa\rangle$ we have
\begin{equation} \langle
x|\cdot\langle\kappa|\psi\rangle\rangle=\langle
x|\cdot\langle\kappa|gh\rangle^x\cdot|s\rangle=\langle\kappa|gh\rangle^x\langle
x|s\rangle\label{cr}
\end{equation}
since $\hat x$ is diagonal.

We observe that
\begin{equation}
[\hat H,\hat Q]=[\hat H,\hat Q^\dagger]=0
\end{equation}
since $[\hat x,\hat Q]=[\hat p,\hat Q]=0$, but $[\hat H,\hat R]\neq 0$ as
$[\hat p,\hat R]\neq 0$ (while $[\hat x,\hat R]=0$).  The $N=2$ supersymmetry
is therefore explicitly broken to $N=1$. In the Appendix we discuss
an $N=2$ supersymmetric extension of the system dynamics.

Consider the subspace in the extended space defined by the conditions
\begin{equation}
\hat Q|ph\rangle\rangle=\hat Q^\dagger|ph\rangle\rangle=0\,.
\label{17}
\end{equation}

In the Fock space representation of the ghost Hilbert space, one can show
that the solutions of (\ref{17}) are
\begin{equation}
|ph\rangle\rangle=\lambda|0\rangle^x_{gh}\cdot |s\rangle
+\sum_{n=1}^\infty|\phi_n\rangle^x_{gh}\cdot|s_n\rangle\,.
\label{18}
\end{equation}
Here $|s\rangle$ and $|s_n\rangle$ are system states and
\begin{equation}
|\phi_n\rangle^x_{gh}=\hat\Gamma^n|0\rangle_{gh}^x=
\hat Q\hat\Phi\hat\Gamma^{n-1}|0\rangle_{gh}^x=
\hat Q^\dagger\hat\Phi^\prime\hat
\Gamma^{n-1}|0\rangle_{gh}^x \label{19}
\end{equation}
with
\begin{equation}
\hat\Gamma=\hat b_1^\dagger \hat b_2^\dagger +\hat c_1^\dagger\hat
c_2^\dagger\;,\quad\hat\Phi=\hat b_1^\dagger\hat c_2^\dagger\;,\quad
\hat\Phi^\prime=\hat b_2^\dagger\hat c_1^\dagger\,.
\end{equation}

Note that $\langle\langle ph|ph\rangle\rangle=0$ if $\lambda=0$ in (\ref{18})
since vectors (\ref{19}) have, by definition (\ref{17}), zero norms.

We remark that the choice of Fock space to discuss the
properties of the states (\ref{17}) is only a matter of convenience.  Indeed,
it is easy to prove that these properties hold generally, i.e., a state
satisfying (\ref{17}) in an arbitrary Hilbert space is the sum of a closed
state (belonging to ${\rm kern}(\hat Q)/{\rm image}(\hat Q)
\cap{\rm kern}(\hat Q^\dagger)/{\rm image}(\hat Q^\dagger)$) and an
exact state (expressible as $\hat Q$ or $\hat Q^\dagger$ on some state).  By
definition (\ref{17}) the latter have zero norm.  In Fock space the ghost
vacuum is the only closed state.

For any operator $\hat O$ commuting with $\hat Q$ and $\hat Q^\dagger$,
it follows from
(\ref{19}), (\ref{17}) and (\ref{18}) that
\begin{equation}
\langle\langle ph^\prime|\hat O|ph\rangle\rangle=\langle
s^\prime|\cdot{}_{gh}^x\langle 0|\hat O|0\rangle_{gh}^x\cdot|s\rangle
\label{20}
\end{equation}
if $\lambda=\exp(i\alpha)$ and $\alpha$ is independent of $x$  (Even in the
case where $\alpha$ is $x$-dependent this relation still holds, provided that
$\hat p$ is replaced by $\hat p-d\alpha/dx$).  For $\hat O=1$ this yields
$\langle\langle ph^\prime|ph\rangle\rangle=\langle s^\prime|s\rangle$.

If $\hat O=\hat O_s=O(\hat x,\hat p)$ equation (\ref{20}) becomes
\begin{equation}
\langle\langle ph^\prime|\hat O_s|ph\rangle\rangle=\langle
s^\prime|\hat O_s|s\rangle\,.
\label{21}
\end{equation}

To prove (\ref{21}) we note that since $\hat x$ commutes with
$|0\rangle_{gh}^x$, it acts directly on the system state.  We only have to
prove
that the same holds for $\hat p$.  From the relations (\ref{7}), (\ref{8}) one
easily verifies $[\hat p,\hat b_1]=\beta \hat b_2^\dagger$, where
$\beta=-i\omega^\prime/(2\omega)$, and similar relations for the other creation
and annihilation operators.  From the definition of the ghost vacuum we have $
[\hat p,\hat b_1|0\rangle^x_{gh}]=\hat b_1[\hat p,|0\rangle^x_{gh}]+\beta
\hat b_2^\dagger|0\rangle^x_{gh}=0$ and analogues relations for the other
annihilation operators.  As $\hat p$ commutes with $\hat Q$ and $\hat
Q^\dagger$, these relations imply that the ghost state $[\hat p,|0
\rangle_{gh}^x]$ belongs
the physical subspace (\ref{19}) with particle number $N_{gh}=2$.
We find from (\ref{19})
$ [\hat p,|0\rangle_{gh}^x]= \hat Q\hat\Phi|0\rangle_{gh}^x\equiv \hat
Q|gh\rangle^x$.  Thus for $\hat O_s =\hat p^n$ we deduce
\begin{equation}
{}_{gh}^x\langle 0|\hat p^n|0\rangle_{gh}^x= {}_{gh}^x\langle 0|\hat
p^{n-1}([\hat p,|0\rangle_{gh}^x]+|0\rangle_{gh}^x \hat p)=
{}_{gh}^x\langle 0|\hat p^{n-1}|0\rangle_{gh}^x\hat p=\hat p^n\,.
\end{equation}

Relations (\ref{21}) as well as (\ref{9}), (\ref{10}) and (\ref{11}) imply
that $\langle\langle ph^\prime|\hat H|ph \rangle\rangle= \langle\langle
ph^\prime|\hat H_s|ph \rangle\rangle=\langle s^\prime|\hat H_s|s\rangle$.
Since both $\hat H$ and $\hat H_s$ commute with $\hat Q$ and $\hat Q^\dagger$,
their application to a physical state leads to a physical state and a simple
inductive argument then shows that $\langle\langle ph^\prime|\hat H^n|ph
\rangle\rangle= \langle\langle ph^\prime|\hat H_s^n|ph \rangle\rangle=\langle
s^\prime|\hat H_s^n|s\rangle$ where (\ref{21}) was used in the latter
equality.  We conclude
\begin{equation}
\langle\langle ph^\prime|e^{-it\hat H}|ph
\rangle\rangle=\langle s^\prime|e^{-it \hat H_s}|s\rangle\,.
\label{25}
\end{equation}

As an application of (\ref{25}) we take $|s\rangle=|x\rangle$ and
$|s^\prime\rangle=|x^\prime\rangle$.  This yields
\begin{equation}
\langle\langle ph^\prime|e^{-it\hat H}|ph
\rangle\rangle=\int [dx]e^{i S_s(x)}\;,
\label{26}
\end{equation}
where $S_s(x)=\int_0^t dt^\prime(\dot x^2/2-V(x))$ is the system action.  The
trajectories contributing in the functional integral obey the boundary
conditions $x(0)=x$ and $x(t)=x^\prime$.

The amplitude in (\ref{26}) can be represented as a functional integral over
the extended configuration space.  Consider the functional integral
representation of the transition amplitude in the extended configuration space
\begin{equation}
\langle\langle x^\prime,\kappa^\prime|e^{-it\hat H}|x,\kappa\rangle\rangle=
\int [dx][dz^*][dz][d\eta^*][d\eta]e^{iS_e}\;,
\label{27}
\end{equation}
where $|x,\kappa\rangle\rangle=|\kappa\rangle\cdot|x\rangle$ and
\begin{equation}
S_e=S_s+\int_0^tdt^\prime(\dot z^*\dot
z+\dot\eta^*\dot\eta-\omega^2(x)(z^*z+\eta^*\eta))\equiv S_s+S_{gh}\,.
\label{2.28}
\end{equation}
The trajectories for the ghost variables obey the boundary conditions
$\kappa(0)=\kappa$ and $\kappa(t)=\kappa^\prime$.

The functional integral (\ref{27}) coincides with (\ref{26}) if we calculate
its convolution with any two physical states (\ref{17}).  There is a particular
choice of boundary conditions for the ghost fields in (\ref{27}) such
that (\ref{27}) coincides with (\ref{26}).  This is achieved by choosing
$\kappa^\prime=\kappa=0$.  This implies that we choose in (\ref{26})
$|ph\rangle\rangle=|0\rangle\cdot|x\rangle$ and
$|ph^\prime\rangle\rangle=|0\rangle\cdot|x^\prime\rangle$ where
$\hat\kappa|0\rangle=0$ (not to be confused with the ghost vacuum).
To verify that these states satisfy (\ref{17}), one
expresses the operators $\hat Q$ and $\hat Q^\dagger$ through relations
(\ref{7}), (\ref{8}) in terms of the ghost canonical variables.  Using
the coordinate representation (\ref{cr1}), (\ref{cr2}) we obtain
\begin{equation}
\langle\kappa|\hat Q|0\rangle =\langle\kappa|\hat Q^\dagger|0\rangle=0
\end{equation}
since $\langle\kappa|0\rangle=\delta(\kappa)$ ($\int
d\kappa\delta(\kappa)\psi(\kappa)=\psi(0)$).

Note that (\ref{27}) is a Gaussian integral for the ghost variables.  Therefore
\begin{equation}
\int[d\kappa]\,e^{iS_{gh}}=F(x,\kappa^\prime,\kappa)\Delta_F/\Delta_B=
F(x,\kappa^\prime,\kappa)
\end{equation}
where $\Delta_F=\Delta_B=\det(-\partial_t^2-\omega^2)$ and $
F(x,\kappa^\prime,\kappa)=\exp(iS_{gh}(\kappa_{cl}))$ with $\kappa_{cl}$ a
classical solution obeying the boundary conditions $\kappa(0)=\kappa$ and
$\kappa(t)=\kappa^\prime$.

Relation (\ref{26}) implies
\begin{equation}
\int d\kappa d\kappa^\prime\;{}_{gh}^x\langle ph^\prime|\kappa^\prime\rangle
F(x,\kappa^\prime,\kappa)\langle \kappa|ph\rangle_{gh}^x=1\,.
\label{bc}
\end{equation}
For vanishing ghost boundary conditions,
$\langle\kappa|ph\rangle_{gh}^x=\delta(\kappa)$ and $F[x,0,0]=1$.

\section{A gauge model.}
\setcounter{equation}0

In this section we illustrate, by means of a simple
model with a $SU(2)$ gauge symmetry, how the method of the previous section
can be used to construct a functional integral for a gauge theory without
fixing a gauge.

Let the canonical variables $\hat p$ and $\hat x$ be elements of a linear
unitary representation of $SU(2)$.  We choose the system Hamiltonian as in
(\ref{sh}), but now $\hat p^2=(\hat p,\hat p)$ where $(\;,\;)$ is the
invariant inner product.  We denote by $\lambda^a$ the $SU(2)$ generators in
the chosen representation.  The Hamiltonian (\ref{sh}) describes a gauge
theory if the physical states are gauge invariant {\cite{dir}}, i.e.,

\begin{equation}
\hat\sigma_s^a|ph\rangle_s=i\hat x\lambda^a\hat p|ph\rangle_s\equiv i(\hat
x,\lambda^a\hat p)|ph\rangle_s=0
\label{3.4}
\end{equation}
and $[\hat H_s,\hat \sigma^a_s]=0$.  The canonical commutation relation
$[\hat x,\hat p]=i$ implies
$[\hat\sigma^a_s,\hat\sigma^b_s]=2i\epsilon^{abc}\hat\sigma^c_s$ so that the
constraints generate $SU(2)$.  The Hamiltonian (\ref{sh}) is
invariant
under gauge transformations generated by $\hat\sigma_s^a$, $\hat p\rightarrow
T_g\hat p$, $\hat x\rightarrow T_g\hat x$ and
$T_gT_g^\dagger=T_g^\dagger T_g=1$.

As described in section 2, we extend the system Hamiltonian (\ref{sh})
by adding a ghost  Hamiltonian $\hat H=\hat H_s+\hat H_{gh}$, where
\begin{equation}
\hat H_{gh}=(\hat p_z^\dagger,\hat p_z) +(\hat p_\eta^\dagger,\hat p_\eta) +
(\hat z^\dagger,\Omega(\hat x)\hat z) +(\hat \eta^\dagger,\Omega(\hat x)
\hat\eta)
\label{ghh}
\end{equation}
and $\Omega$ is a hermitian, strictly positive matrix.
For simplicity we assume the
ghost variables to realize the fundamental representation of $SU(2)$, i.e.,
they are complex isospinors.

To make (\ref{ghh}) invariant under the $SU(2)$ transformations $\hat
z\rightarrow\tilde T_g\hat z$, $\hat \eta\rightarrow\tilde T_g\hat \eta$,
with
$\tilde T_g$ a $2\times 2$ unitary matrix in the fundamental
representation of
$SU(2)$, we require the following transformation rule for
$\Omega$: $\Omega(T_g\hat x)
=\tilde T_g \Omega(\hat x)\tilde T^\dagger_g$ so that $\Omega$
transforms in the adjoint representation.

Introducing the fermionic operators
\begin{equation}
\hat Q=i[(\hat\eta^\dagger,\hat p_z)-(\hat p_\eta^\dagger,\hat z)]\;,\quad
\hat R(\hat x)=i[(\hat p_\eta^\dagger,\hat p_z)+(\hat\eta^\dagger,\Omega(\hat
x)\hat z)]
\end{equation}
and their adjoints $\hat Q^\dagger$, $\hat R^\dagger(\hat x)$ we find
\begin{equation}
\hat H_{gh}=[\hat Q,\hat R^\dagger(\hat x)]=[\hat Q^\dagger,\hat R(\hat
x)]\,.
\end{equation}
Note that the operators $\hat Q$, $\hat Q^\dagger$ commute with the total
Hamiltonian, while $\hat R(\hat x)$ and $\hat R^\dagger(\hat x)$ commute only
with the ghost Hamiltonian.

As before the Hilbert space of the ghost sector can be build as a Fock space.
Indeed, if $\omega_i^2(\hat x)$ are the eigenvalues of $\Omega(\hat x)$, then
$\hat
Q=\sum_i\hat Q_i$ and $\hat R(\hat x)=\sum_i\omega_i(\hat x)\hat R_i$ with
$\hat Q_i$ and $\hat R_i$ determined by (\ref{ssc}) for each SUSY oscillator
labelled by $i=1,2$.

The constraints are also extended
$\hat\sigma^a=\hat\sigma_s^a+\hat\sigma^a_{gh}$ with
\begin{equation}
\hat\sigma_{gh}^a=i\hat z^\dagger\tau^a\hat p_z -i\hat p_z^\dagger\tau^a\hat z+
i\hat \eta^\dagger\tau^a\hat p_\eta-i\hat p_\eta^\dagger\tau^a\hat \eta=
[\hat Q,i\hat z^\dagger\tau^a\hat p_\eta-i\hat p_z^\dagger\tau^a\hat \eta]\,.
\label{3.2}
\end{equation}
Here $\tau^a$ are the Pauli matrices.  By construction $[\hat\sigma^a,\hat
H]=0$, $[\hat\sigma^a,\hat Q]= [\hat\sigma^a,\hat Q^\dagger]=0$ and
$\hat\sigma^a$ generate $SU(2)$.

We prove the relation
\begin{equation}
\langle\langle ph^\prime|e^{-it\hat H}|ph\rangle\rangle=
{}_s\langle ph^\prime|e^{-it\hat H_s}|ph\rangle_s\;,
\label{3.3}
\end{equation}
where the physical states $|ph\rangle\rangle$ and $|ph^\prime\rangle\rangle$
in the extended space have non-zero norm and satisfy the conditions
\begin{equation}
\hat Q|ph\rangle\rangle =\hat Q^\dagger|ph\rangle\rangle =
\hat \sigma^a|ph\rangle\rangle=0,
\label{3.5}
\end{equation}
while the system states on the right of (\ref{3.3}) obey (\ref{3.4}).  Relation
(\ref{3.3}) establishes a one-to-one correspondence between physical transition
matrix elements of the original gauge theory and its supersymmetric extension.

From (\ref{25}) we note that it is sufficient to prove that (\ref{3.5}) imply
(\ref{3.4}).  Consider the second order Casimir operator $\hat
C=\hat\sigma^a\hat\sigma^a$.  From (\ref{3.2}) it is easily established that
$\hat C=\hat C_s+[\hat Q,\hat \Lambda]$ where $\hat C_s$ is the second order
Casimir of the system, and the explicit form of $\hat\Lambda$ can be read of
from (\ref{3.2}).  It follows
\begin{equation}
\langle\langle ph|\hat C|ph\rangle\rangle=\langle\langle
ph|\hat C_s|ph\rangle\rangle= \langle s|\hat C_s|s\rangle\;,
\end{equation}
where (\ref{21}) was used.  Since the system Hilbert space has a strictly
positive norm, (\ref{3.5}) implies (\ref{3.4}).
Thus relation (\ref{3.3}) is proved.

Next we turn to the functional integral representation of (\ref{3.3}).  To
avoid divergencies in the measure of the functional integral caused by the
gauge invariance, we have to solve the constraints
$\hat\sigma^a|ph\rangle\rangle=0$ explicitly by choosing a parameterization
(coordinates) in the physical configuration space, i.e., we should fix a gauge
in the extended theory.  In contrast with the BRST-BFV scheme we have boson
ghosts so that we can impose the gauge condition on them, while the system
degrees of freedom remain free of any gauge condition.

To fulfil this program, we choose the coordinate representation (\ref{cr})
$\langle
x|\cdot\langle\kappa|ph\rangle\rangle$
$ =\psi_{ph}(x,\kappa)$,
where all states are
functions of $x$ and $\kappa$.  To project the total Hamiltonian on the
subspace of gauge invariant functions $\psi_{ph}(x,\kappa)$, we change
variables in the superspace \begin{equation}
z=\tilde T_g\chi\rho/\sqrt{2}\,,\quad x\rightarrow T_gx\,,\quad\eta\rightarrow
\tilde T_g\eta\;,
\label{3.6}
\end{equation}
where $\chi^\dagger=(1,0)$, $\rho$ is real and $T_g$ is the group element
$\tilde T_g$ in the representation of $x$.  Denoting the group parameters by
$\theta^a$, the constraint operators in the new coordinates assume the form
$\hat\sigma^a\sim \partial/\partial\theta^a$ {\cite{sha1}}.  Therefore
physical states (gauge invariant states) are functions independent of
$\theta^a$.  We transform $\hat H$ to the curvilinear supercoordinates
(\ref{3.6}) and drop all terms containing $\partial/\partial\theta^a$
to obtain the physical Hamiltonian
\begin{equation}
\hat H_{ph}=\hat H_s+\frac{\hat p^2_\rho}{2}+\frac{3}{8\rho^2}+\hat
p_\eta^\dagger
\hat p_\eta+\frac{1}{2\rho^2}(\tilde\sigma^a)^2+\frac{\rho^2}{2}
\chi^\dagger \Omega\chi+\eta^\dagger \Omega\eta\;,
\label{3.7}
\end{equation}
where $\tilde\sigma^a=\hat\sigma^a_s+\hat\sigma_f^a$ with $
\hat\sigma_f^a=i\eta^\dagger\tau^a\hat p_\eta-i\hat
p_\eta^\dagger\tau^a
\eta$ and $\hat p_\rho=-i\rho^{-3/2}\partial_\rho\circ\rho^{3/2}$
is the hermitian
momentum conjugated to $\rho$.  The third term in (\ref{3.7})
is due to the operator ordering.
Restoring Planck's constant, this would be proportional to $\hbar^2$.

The above procedure to obtain the physical Hamiltonian also applies to the
operators $\hat Q$ and $\hat Q^\dagger$
to derive the supersymmetry generators
on the space of functions invariant under the gauge
transformations generated by $\hat
\sigma^a$.  We find
\begin{eqnarray}
\hat Q_\rho &=&\frac{1}{\sqrt{2}}\eta^\dagger
\chi\partial_\rho-\frac{i\rho}{\sqrt{2}}\hat
p_\eta^\dagger\chi-\frac{1}{\sqrt{2}\rho}
[\eta^\dagger\chi\tilde\sigma^3+\eta^\dagger \psi(\tilde\sigma^2-
i
\tilde\sigma^1)]\,,\nonumber\\
\hat Q_\rho^\dagger&=&-
\frac{1}{\sqrt{2}}\chi^\dagger\eta\partial_\rho
+\frac{i\rho}{\sqrt{2}}\chi^\dagger \hat p_\eta-
\frac{1}{\sqrt{2}\rho}[\chi^\dagger\eta \tilde\sigma^3
+\psi^\dagger\eta(\tilde\sigma^2+i
\tilde\sigma^1)]\label{3.7a}
\end{eqnarray}
with $\psi^\dagger=(0,1)$.
Taking into account that $\eta$ and $\eta^\dagger$ do
not commute with $\tilde\sigma^a$, one can verify
that these two operators are adjoints with respect to the
measure $\rho^3 d\rho$.   Indeed, the term proportional to $1/\rho$,
which appears when the adjoint of
$\partial_\rho$ is taken with respect to this measure, precisely
cancels against a similar term originating
from the reordering of $\eta$ or $\eta^\dagger$ and $\tilde\sigma^a$ in
the adjoint of the third term.  It
can also be checked explicitly that these
operators commute with the physical Hamiltonian (\ref{3.7}) as they should.

The physical states of the system are obtained as before by imposing
\begin{equation}
\hat Q_\rho|ph\rangle\rangle=\hat
Q^\dagger_\rho|ph\rangle\rangle=0\,.\label{3.7b} \end{equation}
We note that the system variables $\hat x$ and $\hat p$ do not commute
with the supersymmetry
generators $\hat Q_\rho$ and $\hat Q^\dagger_\rho$.  Indeed,
the operator $\xi^\dagger\hat Q_\rho +\hat
Q_\rho^\dagger\xi$, with $\xi$ and $\xi^\dagger$ Grassmann variables,
involves a linear combination of
the system constraint operators $\sigma_s^a$.
Therefore it generates an isotopic rotation of  $\hat x$ and
$\hat p$ of the same form as an infinitesimal gauge transformation,
but with the parameters being even
elements of a Grassmann algebra.  Therefore the condition (\ref{3.7b})
ensures gauge invariance in the
vacuum sector  of the ghosts, i.e., a vector describing any system
excitation will automatically satisfy the
Dirac condition (\ref{3.4}).

Note how the reduction in the degrees of freedom is obtained: we start with
$n$ system degrees of
freedom,  $n$ being the real dimension of the representation in which
the $\hat x$ transforms.  Then we
add 4
real bosonic and 4 real fermionic ghost degrees of freedom in a
supersymmetric combination. Since the
system variables realize a trivial representation of
the supersymmetry an exact
cancellation of the ghost degrees of freedom is ensured,
leaving  only the $n$ system degrees of freedom.
Next we eliminate three real
bosonic ghost degrees of freedom by projecting on the
subspace of gauge invariant functions (see
(\ref{3.6})).  The projection (\ref{3.6}) mixes the system
and ghost degrees of freedom in a non-linear way.
In particular the angular variables of the bosonic ghosts
are absorbed in the new system variables so that
three angular variables of $x$ are the supersymmetric partners
of three fermionic ghost degrees of
freedom.  The condition (\ref{3.7b}) ensures that the 4 fermionic
ghost degrees of
freedom are balanced by the $\rho$ ghost degree of freedom and 3
system degrees of freedom.   Hence, we
are
left with $n-3$ degrees of freedom, which is correct as we have
3 independent constraints (\ref{3.4}) in
the original model.

In our arguments above we have assumed that the system constraints
$\hat\sigma^a_s$ are irreducible,
i.e., they are all independent, which would not be the case if,
for example, we take $x$ to be in the adjoint
representation.  Nonetheless, in our approach the supersymmetry
condition (\ref{3.7b}) will still provide
the elimination of the right number of unphysical degrees of
freedom in the system sector, provided the
ghost sector is chosen so that the constraints $\hat\sigma_{gh}^a$
(\ref{3.2}) are irreducible.  The
latter is always possible so that, in contrast to the conventional
BRST-BFV formalism, we do not need
ghosts for ghosts if the system exhibits reducible constraints.

Returning to the functional integral representation,
let $\kappa_\rho$ denote a set
$(\rho,\eta^\dagger,\eta)$.
Using the formalism
of {\cite{sha1}} we find
\begin{equation}
\langle\langle x^\prime,\kappa_\rho^\prime|e^{-it\hat
H_{ph}}|x,\kappa_\rho\rangle\rangle=\int\limits_{-\infty}^\infty
d\kappa_\rho^{\prime\prime}dx^{\prime\prime}\,(\rho^3\rho^{\prime\prime
3})^{-1/2} U_t^{eff}(x',\kappa_\rho',
x^{\prime\prime},\kappa_\rho^{\prime\prime})
Q_S(x^{\prime\prime},\kappa_\rho^{\prime\prime},x,\kappa_\rho)\;;
\label{3.11}
\end{equation}
the amplitude $U^{eff}_t$ is given by a functional integral for the
extended theory in the unitary gauge $z=\chi\rho$:
\begin{eqnarray}
U_t^{eff}(x,\kappa_\rho,x^{\prime\prime},\kappa_\rho^{\prime\prime})&=&
\int [d p][dx][d\kappa_\rho][dp_{\kappa_\rho}]\,e^{iS_H^{ph}}\label{3.8}\\
&=&\int [dx][dy][d\kappa_\rho][\Pi_{t=0}^t|\rho|^3]\,e^{iS^{ph}}\,.\label{3.9}
\end{eqnarray}
Here
\begin{equation}
S_H^{ph}=\int_0^tdt^\prime(\dot x p+p_\eta^\dagger\dot\eta+\dot\eta^\dagger
p_\eta+p_\rho\dot\rho-H_{ph})
\end{equation}
and
\begin{eqnarray}
S^{ph}=\int_0^tdt^\prime\!\!\!&[&\frac{1}{2}(D_tx)^2 + (\tilde
D_t\eta)^\dagger(\tilde
D_t\eta) +\frac{1}{2}\dot\rho^2+\frac{1}{2}\rho^2|y\chi|^2\nonumber\\ &&-V(x)-
\frac{1}{2}\rho^2\chi^\dagger \Omega\chi-\eta^\dagger \Omega\eta-\frac{3}
{8\rho^2}]\,.
\label{3.10}
\end{eqnarray}
We denote by $D_t=\partial_t+iy^a\lambda^a$ and $\tilde D_t$ the covariant
derivatives in the
representation of $x$ and the fundamental representation
($\lambda_a\rightarrow\tau_a$), respectively.

To go from (\ref{3.8}) to (\ref{3.9}) we first replace the measure
$[d\kappa_\rho][dp_{\kappa_\rho}]$ by the Faddeev-Popov measure
in the unitary
gauge $[d\kappa][dp_\kappa]\Pi_t\Delta_{FP}\delta(z-\chi\rho)\delta(\sigma^a)$.
The Faddeev-Popov determinant for the unitary gauge $z=\rho\chi$ is
$\Delta_{FP}\sim |\rho|^3$.  Correspondingly $S_H^{ph}$ in (\ref{3.8}) is
replaced by the Hamiltonian action for the total extended Hamiltonian.  Next
the identity $\int[dy^a]\exp(iy^a\sigma^a)=\Pi_t\delta(\sigma^a)$ is used and
the integrals over all momenta are done.  Finally the integral over $z$ and
$z^\dagger$ are done using the delta function and (\ref{3.9})
is obtained.  Note that apart from the operator ordering
term ($\sim \hbar^2/\rho^2$) in
(\ref{3.10}) all the other terms depending on $\rho$ can be written as
$|\tilde D_tz|^2-z^\dagger \Omega z$ where $z=\rho\chi$.  Thus (\ref{3.10}) is
the
action of the extended supersymmetric gauge theory in the unitary gauge imposed
on the boson ghost $z$.

The kernel $Q_S$ in (\ref{3.11}) takes care of the global properties of the
change of variables (\ref{3.6}) {\cite{sha1,prok1}}
\begin{equation}
Q_S(x^{\prime\prime},\kappa_\rho^{\prime\prime}x^\prime,\kappa_\rho^\prime)=
\sum_s\delta(x^{\prime\prime}-\hat
sx^\prime)\delta_{gh}(\kappa_\rho^{\prime\prime},\hat s\kappa_\rho^\prime)
\label{3.12}
\end{equation}
($\int
d\kappa^\prime\delta_{gh}(\kappa^\prime,\kappa)\psi(\kappa^\prime)
=\psi(\kappa)$).
The sum is over all inequivalent elements $\hat s$ acting on
$x$ and $\kappa_\rho$ as follows: $\hat sx=T_sx$ and $\hat s
\kappa_\rho=\tilde
T_s\kappa_\rho$ where, by definition, $\tilde T_s\chi=\pm\chi$ and $T_s$ are
corresponding $SU(2)$ elements in the representation of $x$.  The $\hat
s$-transformations form a $Z_2$ subgroup of $SU(2)$, $\tilde T_g=\pm 1$.

We note that any physical state $\psi(x_\rho,\kappa_\rho)$ is invariant under
the $\hat s$-trans\-for\-ma\-tions.
Here we denote the transformed system variables
defined in (\ref{3.6}) by $x_\rho$ to distinguish from the original system
variables.  We also define the action of $\hat s$ on $\theta^a$ by $\tilde
T_g(\hat s\theta)=\tilde T_g(\theta)\tilde T_s^{-1}$.  Then $\hat
sx=x$ and $\hat s\kappa=\kappa$ so that $\psi(\hat sx,\hat
s\kappa)=\psi(x,\kappa)$ for any element of the extended Hilbert space.
For a
physical state we have $\psi_{ph}(x,\kappa)=\psi_{ph}(T_g(\theta)x_\rho,\tilde
T_g(\theta)\kappa_\rho)=\psi_{ph}(x_\rho,\kappa_\rho)$.  Since $\hat s$ leaves
$x$ and $\kappa$ invariant, we can replace $\theta^a$, $x_\rho$ and
$\kappa_\rho$ by their $\hat s$-transforms in the latter equality.  Thus we
conclude
$\psi_{ph}(\hat sx_\rho,\hat s\kappa_\rho)= \psi_{ph}( x_\rho,\kappa_\rho)$.
The time evolution must respect this symmetry.  However, the functional
integrals (\ref{3.8}) or (\ref{3.9}) do not respect this symmetry so that they
cannot describe the correct time evolution.  The particular linear combination
of functional integrals (\ref{3.9}) with different boundary conditions as
prescribed by (\ref{3.11}) and (\ref{3.12}) has the desired invariance
property {\cite{sha1,prok1}}.

From (\ref{3.3}) we know that (\ref{3.11}) coincides with the physical system
transition amplitude only after supersymmetric and gauge invariant boundary
conditions have been imposed on the ghosts.  In analogy with section 2, the
state $\langle\langle x,\kappa|\psi\rangle\rangle=\delta_{gh}(\kappa)
\psi_s(x)$
is supersymmetric, i.e., it is annihilated by both supersymmetric generators.
However, it is not gauge invariant so that it cannot be used as boundary
conditions in (\ref{3.11}).  To construct a gauge invariant supersymmetric
state we average $\delta_{gh}(\kappa)\psi_s(x)$  over the gauge
transformations.  This yields
\begin{equation}
\langle\langle\kappa_\rho,x|ph\rangle\rangle=\frac{2}{|\rho|^3}\delta(\rho)
(\eta^\dagger\eta)^2\psi_s^{ph}(x)= \frac{2}{z^\dagger z}\delta(z^\dagger z)
(\eta^\dagger\eta)^2\psi_s^{ph}(x)\;,
\label{3.13}
\end{equation}
where $\psi_s^{ph}(x)$ is any gauge invariant function of $x$.  The ghost part
of (\ref{3.13}) is just a delta function with respect to the scalar product
measure $\int_0^\infty d\rho\rho^3\int d\eta^\dagger d\eta$.  Substituting
(\ref{3.13}) into (\ref{3.3}) and using (\ref{3.11}) we obtain
\begin{equation}
{}_s\langle ph^\prime|e^{-it\hat H_s}|ph\rangle_s=\int dx dx^\prime \langle
ph^\prime|x^\prime\rangle \tilde U^{eff}_t(x^\prime,x)\langle x|ph\rangle
\label{3.15}
\end{equation}
with
\begin{equation}
\tilde U^{eff}_t(x^\prime,x)=\lim_{\kappa_\rho,\kappa^\prime_\rho\rightarrow 0}
(\rho\rho^\prime)^ {-3/2}
U^{eff}_t(x^\prime,\kappa_\rho^\prime,x,\kappa_\rho)\,. \label{3.14}
\end{equation}

There is no problem to impose zero boundary conditions on the fermion ghosts in
the functional integral (\ref{3.9}).  For the boson ghost $\rho$ one can only
prove that the limit (\ref{3.14}) exists.  Indeed, in the neighborhood of
$\rho=0$ or $\rho^\prime=0$, the integral (\ref{3.9}) satisfies the
Schr\"odinger equation $(-\partial_\rho^2/2+3/(8\rho^2))U^{eff}_t=0+O(1)$.
Therefore $U_t^{eff}\sim(\rho\rho^\prime)^{3/2}$ as $\rho$ and $\rho^\prime$
approach zero and the limit (\ref{3.14}) exists.

\section{Yang-Mills theories.}
\setcounter{equation}0

Here we briefly outline how the procedure of the
previous sections can be generalized to Yang-Mills theories.  A more detailed
discussion can be found in {\cite{sch}}.

One can go from the gauge model of section 3 to the Yang-Mills theory (with
$SU(2)$ gauge symmetry) by simple analogy.  We identify the vector gauge
potentials $A_i\,,\,(i=1,2,3)$ with the system variables $x$, the Lagrange
multiplier $y$ with $A_0$ and $\kappa$ becomes a set of ghost fields.  One can
convince oneself that the expressions
(\ref{3.15}) and (\ref{3.14}) still hold with the replacements $x\rightarrow
A_i$ and $dx\rightarrow \Pi_x dA_i(x)$, where the product is taken over all
space points.  Similarly the measure $\rho^3$ is replaced by the functional
measure $\mu=\Pi_x\rho^3(x)$ and
\begin{equation}
U^{eff}_t[A',\kappa_\rho',A,\kappa_\rho]=\int[dA_\mu][d\rho]
\mu[\rho][d\eta^\dagger][d\eta]\,e^{iS^{eff}}
\label{3.16}
\end{equation}
with
\begin{equation}
S^{eff}=\int d^4x[-\frac{1}{4}F_{\mu\nu}^2+(\tilde D_\mu\eta)^\dagger(\tilde
D^\mu\eta)
+\frac{1}{2}(\partial_\mu\rho)^2+\frac{1}{2}A_\mu^2\rho^2-m^2(
\frac{\rho^2}{2}+\eta^\dagger\eta)-
V_q(\rho)]\,.
\label{3.17}
\end{equation}
Here $\tilde D_\mu$ is the covariant derivative in the fundamental
representation and a specific choice of
the matrix $\Omega$ (see (\ref{ghh})) was
made, $\Omega= -\tilde{D}_i^2(A)+m^2$,
(in order for $\Omega$ to be strictly positive a mass term is
added for the ghosts).  The quantum
corrections to the potential energy is
proportional to a divergent factor $V_q(\rho)
=3[\delta^3(0)]^2/(8\rho^2)$ as is
usual for operator ordering terms in a field theory.  Making use of a lattice
regularization of space in (\ref{3.17}) our earlier arguments can be repeated
to prove the existence of the limit (\ref{3.14}) for each degree of freedom of
the field $\rho(x)$.  The coupling of the field oscillators due to the term
$(\partial_i\rho)^2$ is of no consequence in the limit $\rho\rightarrow 0$ as
it is negligible in comparison with the singular quantum potential $V_q$.
Therefore the limit (\ref{3.14}) exists and supersymmetric gauge invariant
boundary conditions can be imposed as before.

Due to the presence of the singular potential $V_q(\rho)$ the limit
(\ref{3.14}) is highly undesirable and
leads to technical difficulties.
This limit originates from our choice of  supersymmetric
boundary conditions for the path integral. Fortunately, all physically
relevant information about a quantum system can always be extracted
from Green's functions. In the next section we show that
the restriction of the supersymmetric boundary conditions can be
dropped in the path integral representation of Green's function and,
hence, the aforementioned technical difficulty can be avoided.

\section{System Green's functions.}
\setcounter{equation}0

The quantities one is normally interested
in are the Green's functions of the system from which the
S-matrix elements and the excitation spectrum can be extracted.
They are defined as the vacuum (groundstate) expectation values
of time ordered products of
the field operators.  For complex time they are
defined by the following quantities,
taken in the limit $\tau_f\rightarrow\infty$, $\tau_i\rightarrow -
\infty$:
\begin{equation}
\frac{\langle s^\prime|e^{-\hat H_s \tau_f}\hat O^s_1(\tau_1)
\hat O^s_2(\tau_2)\ldots \hat O^s_n(\tau_n)
e^{\hat H_s
\tau_i}|
s\rangle}{\langle s^\prime|e^{-\hat H_s (\tau_f-\tau_i)}|s\rangle}\,.
\label{5.1}
\end{equation}
Here $\tau_f>\tau_1>\ldots\tau_n>\tau_i$, $|s\rangle$,
$|s^\prime\rangle$ are arbitrary
system states and
$\hat O^s_i(\tau_i)= e^{\hat H_s \tau_i}\hat O^s_i e^{-\hat H_s \tau_i}$ are
system operators in the Heisenberg picture.
The real time Green's functions are obtained from the complex time Green's
functions by analytic
continuation (Wick rotation). Following the standard procedure one
can also express the
Green's functions as functional integrals.

When the system possesses a gauge symmetry, it is again necessary to
eliminate the unphysical degrees of
freedom to obtain well defined Green's functions.   Following the
Faddeev-Popov approach, one is
again faced by the difficulties associated with gauge fixing.
Therefore we would like to extend our
procedure to the evaluation of system Green's functions.
In principle this is
straightforward, except for
the following point which require special care:
The groundstate of the supersymmetric extended
Hamiltonian may turn out not to be supersymmetric, i.e., the supersymmetry
is spontaneously broken and
the groundstate does not satisfy (\ref {17}) and is thus unphysical.
If this happens the
Green's functions
of the supersymmetric extended system do not coincide with those of
the original system, as the
expectation value of the system operators are evaluated with respect
to an unphysical state.  One way of
avoiding this difficulty is to impose supersymmetric boundary conditions
in the functional integral for the
Green's functions as relation (\ref{25}) holds regardless whether
the supersymmetry is spontaneously
broken or not.  From our previous discussion it is, however, clear that these
boundary conditions are inconvenient from a calculational point of view.
Instead we present here an
alternative way of obtaining the system Green's function which allows us to
choose  arbitrary boundary conditions.

For the purpose of illustration let us consider again the 1-dimensional
quantum system of section 2.  The
complex time Green's functions of the supersymmetric extended system is
obtained from a simple generalization of (\ref{5.1})
\begin{equation}
\frac{\langle \langle \psi^\prime|e^{-\hat H \tau_f}
\hat O_1(\tau_1)\hat O_2(\tau_2)\ldots \hat O_n(\tau_n)
e^{\hat H
\tau_i}|
\psi\rangle\rangle}{\langle \langle \psi^\prime|e^{-\hat H
(\tau_f-\tau_i)}|\psi\rangle\rangle}\,.
\label{5.1a}
\end{equation}
Here $|\psi\rangle\rangle=|gh\rangle\cdot|s\rangle$,
$|\psi^\prime\rangle\rangle=|gh^\prime\rangle\cdot|s^\prime\rangle$
are states in the extended
Hilbert space and $\hat H$ is the supersymmetric extended Hamiltonian
\begin{equation}
\hat H=\hat H_s+\alpha\hat H_{gh}\label{5.2a}
\end{equation}
where $\hat H_s$, $\hat H_{gh}$ were defined in  (\ref{sh},\ref{2})
and $\alpha$ is an arbitrary real
number.  The operators $\hat O_i(\tau_i)=
e^{\hat H \tau_i}\hat O_i e^{-\hat H
\tau_i}$ are supersymmetric
extensions of the system operators in the Heisenberg picture defined by
$\hat O_i= \hat O^s_i+[\hat
Q,(\hat O_i)_{gh}]$ with any desired choice of $(\hat O_i)_{gh}$.

 Note that the extended  Hamiltonian above differs slightly from that
 of (\ref{11})  in the presence of the
arbitrary real number $\alpha$.  Since the ghost Hamiltonian is irrelevant
on the physical subspace (see
(\ref{21},\ref{25})),
physical quantities will be independent of $\alpha$, while excitations in
the ghost sector will be affected.
We now
exploit this fact to our advantage by considering the limit
$\alpha\rightarrow\infty$.  From eq. (\ref{9})
we note that  this limit will
push all states with non-zero ghost number to infinite energy,
while the states with zero ghost
number will be untouched.  Thus in this limit the
groundstate of the extended Hamiltonian will
be supersymmetric and assumes the form
$|0\rangle^x_{gh}\cdot|gs\rangle_s$,
where $|gs\rangle_s$ is the system groundstate
(see (\ref{18})), so that the Green's
functions of the extended Hamiltonian coincide
with those of the system Hamiltonian.  We now prove
that this is indeed the case.

Let $|N,x\rangle\rangle =|N\rangle^x_{gh}\cdot|x\rangle$
be eigenstates of $\hat
N_{gh}$ and $\hat x$ where
we noted that $[\hat N_{gh},\hat x]=0$.
The states $|N,x\rangle\rangle$ form a complete basis in the
extended Hilbert space ($|N\rangle^x_{gh} $ is the Fock basis of the ghosts).
Writing out the Heisenberg
operators in (\ref{5.1a}) and inserting a complete set of states in the form
\begin{equation}
1=\sum_N\int dx|N,x\rangle\rangle\langle \langle x, N| \label{5.2}
\end{equation}
before and after each time evolution operator, the Green's function
(\ref{5.1a}) reduces to the product of generic
matrix elements of the extended system operators
$\langle \langle x^\prime,
N^\prime,|\hat O_i$ $|N,x\rangle\rangle$  and the  time
evolution operator $\langle \langle x^\prime, N^\prime|e^{-T(\hat H_s+\alpha
\hat H_{gh})}|N,x\rangle\rangle\equiv
\langle \langle x^\prime, N^\prime|\hat U_T$ $|N,x\rangle\rangle$ where
$T=\tau_j-\tau_{j+1}>0$.
Consider the latter matrix elements.  From
\begin{equation}
 \hat H_{gh}|N,x\rangle\rangle=\omega(x)N|N,x\rangle\rangle\label{5.3}
\end{equation}
follows that in the limit $\alpha\rightarrow\infty$ a perturbation expansion
in $\hat H_s$ can be
developed, using $\alpha\hat H_{gh}$ as the unperturbed Hamiltonian.
Introducing an interaction picture
we can express the time evolution operator as:
\begin{eqnarray}
\hat U_T&=& e^{-\alpha\hat H_{gh}T}\sum_{n=0}^\infty
(-1)^n\int\limits^T_0 d\tau_1\int \limits^{\tau_1}_0 d\tau_2
\ldots\int\limits
^{\tau_{n-1}}_0d\tau_n
\hat H_s(\tau_1) \hat H_s(\tau_2)\ldots \hat H_s(\tau_n)\nonumber\\
&\equiv& e^{-\alpha\hat H_{gh}T}\sum_n(-1)^n\hat B_n\,,\label{5.4}
\end{eqnarray}
where $T>\tau_1>\tau_2\ldots\tau_n>0$ and $\hat H_s(\tau_i)= e^{\alpha\hat
H_{gh} \tau_i}\hat H_s e^{-
\alpha \hat H_{gh} \tau_i}$.  To evaluate the matrix elements
$\langle \langle x^\prime, N^\prime|\hat U_T|N,x\rangle\rangle$,
consider a typical
term in the series (\ref{5.4}) and insert the complete set of states
(\ref{5.2}) between the
$\hat H_s(\tau_i)$.  Using (\ref{5.3}) we find:
\begin{eqnarray}
&&\langle \langle x^\prime, N^\prime |\hat B_n|N,x\rangle\rangle \nonumber\\
&&=\sum_{N_1\ldots N_{n-1}}\!\!\!\int dx_1\ldots dx_{n-1}
\int \limits^T_0 d\tau_1\int\limits^{\tau_1}_0 d\tau_2\ldots
\int\limits^{\tau_{n-1}}_0
d\tau_n
e^{-\alpha\omega^\prime N^\prime(T -
\tau_1)}\langle \langle x^\prime,N^\prime |
\hat H_s|N_1,x_1\rangle\rangle \nonumber\\
&&\times e^{-\alpha\omega_1
N_1(\tau_1-
\tau_2)}\langle \langle x_1, N_1| \hat H_s |N_2,x_2\rangle\rangle\ldots
e^{-\alpha\omega_{n-1} N_{n-1}(\tau_{n-1}-
\tau_n)}\nonumber\\
&&\times\langle \langle x_{n-1},N_{n-1} |\hat H_s|N,x\rangle\rangle
e^{-\alpha\omega
N\tau_n}\,.\label{5.5}
\end{eqnarray}
Here the shorthand notation $\omega^\prime=\omega(x^\prime)$,
$\omega_i=\omega(x_i)$ and
$\omega=\omega(x)$ was used.
Since $T>\tau_1>\tau_2\ldots\tau_n>0$ we deduce that in the limit
$\alpha\rightarrow\infty$ only terms with $N^\prime=N_1=\ldots=N=0$ survive.
The ghost vacuum
($N=0$) is a physical state and therefore
$\langle\langle x^\prime,0|\hat H_s|0,x\rangle\rangle =\langle
x^\prime|\hat H_s|x\rangle$ according to (\ref{21}).  Making use of
these observations we find that eq. (\ref{5.5}) becomes
\begin{equation}
\langle \langle x^\prime, N^\prime |\hat B_n|N,x\rangle\rangle
\rightarrow \frac{T^n}{n!}\langle x^\prime|\hat
H_s^n|x\rangle\delta_{N^\prime,0}\delta_{N,0} \,, \label{5.6}
\end{equation}
when $\alpha\rightarrow\infty$.   From (\ref{5.6}) we conclude
\begin{equation}
\langle \langle x^\prime,N^\prime |\hat U_T|N,x\rangle\rangle\;
\rightarrow\; \langle x^\prime|e^{-\hat H_s
T}|x\rangle\delta_{N^\prime,0}\delta_{N,0}\,,
\quad\alpha\rightarrow\infty\,.\label{5.7}
\end{equation}

Substituting this result in (\ref{5.1a}) and using (\ref{21}),
we obtain in the limit
$\alpha\rightarrow\infty$
\begin{equation}
\frac{\langle \langle \psi^\prime|e^{-\hat H \tau_f}
\hat O_1(\tau_1)
\cdots \hat O_n(\tau_n)
e^{\hat H \tau_i}|
\psi\rangle\rangle}{\langle \langle \psi^\prime|
e^{-\hat H (\tau_f-\tau_i)}|\psi\rangle\rangle}
\rightarrow \frac{\langle s^\prime|e^{-\hat H_s \tau_f}
\hat O^s_1(\tau_1)
\cdots \hat O^s_n(\tau_n) e^{\hat H_s
\tau_i}|
s\rangle}{\langle s^\prime|
e^{-\hat H_s (\tau_f-\tau_i)}|s\rangle}\,. \label{5.8}
\end{equation}

This establishes that by extending the Hamiltonian as in (\ref{5.2a}),
and taking the limit
$\alpha\rightarrow\infty$ before the limits $\tau_f\rightarrow\infty$,
$\tau_i\rightarrow -\infty$, we indeed
recover the system Green's functions from the Green's functions of the
supersymmetric extended system.
In this approach arbitrary boundary conditions can be imposed in
the path integral for the
Green's function as the limit $\alpha\rightarrow\infty$
ensures that the supersymmetric groundstate is selected in the evaluation
of the expectation value.

We conclude this section with a couple of general remarks.
The first remark concerns degeneracies.  We
note that for some values of $x$, $x^\prime$, $N$ and $N^\prime$
the following equality holds:
$\omega(x)N=\omega(x^\prime)N^\prime$ and we have to analyze what
happens to the perturbation
expansion (\ref{5.5}) when this degeneracy occurs in the ghost energy levels
(see (\ref{5.3})).   This
degeneracy does not
affect our conclusions (\ref{5.6}) and, hence, (\ref{5.8}).  Indeed,
the operator $\hat H_s$ is local and
therefore its matrix
elements $\langle x^\prime|\hat H_s|x\rangle$ have support only on
$x=x^\prime$. Consequently this
equality only holds when $N=N^\prime$ and it is necessary  to consider
the degeneracy of a ghost level
with a fixed ghost number only.  In this case some of the exponential
factors disappear and the
corresponding integration yields
a typical factor $\tau_i g_N$, with $g_N$ the degeneracy of the state
$|N\rangle _{gh} $.  Since this
is a finite factor, independent of $\alpha$, we conclude that in the limit
$\alpha\rightarrow\infty$  the
factor  $ e^{-\alpha\omega(x^\prime) N^\prime T}$ ensures that only
contributions from the ghost vacuum survive.

The above argument will hold for any supersymmetric extension of the
Hamiltonian, provided that the ghost Hamiltonian itself does not  exhibit
spontaneous breaking of the
supersymmetry.

Clearly the above procedure would not be necessary if one can prove that the
supersymmetry is not spontaneously broken for $\alpha=1$.
If the groundstate of the extended
Hamiltonian has the form (\ref{18}), it is not hard to see that the Green's
functions (\ref{5.1a}) coincide
with the system Green's functions (\ref{5.1}) in the limit
$\tau_f\rightarrow\infty$, $\tau_i\rightarrow-
\infty$.   However, a proof that the supersymmetry is not spontaneously
broken
does not seem easy.

We remark that the procedure as outlined above does not apply to relativistic
systems as it would break
Lorentz invariance.  However, a relativistic invariant
formulation of the above procedure can be obtained
by taking the limit where the mass of the ghost fields go to infinity,
$m\rightarrow\infty$ (see
(\ref{3.17})).
In this limit contributions of ghost excitations to the Green's
functions are exponentially suppressed
because their energies grow as $m$.  It is then
not difficult to generalize the
above argument to show that our conclusion (\ref{5.8}) holds in this case.

In the case of a gauge system, one is interested in the Green's functions of
gauge invariant system
operators.   They are obtained from our prescription above by considering
only physical operators, $\hat
O_i(\tau_i)$,  in (\ref{5.1a}), i.e.,  operators that commute with the
supersymmetry generators obtained
after projecting onto the space of gauge invariant functions.
For the gauge model of section 3 we would
therefore consider operators  that commute with
the supersymmetry generators $\hat Q_\rho$ and $\hat
Q_\rho^\dagger$ of (\ref{3.7a}).  From the discussion below (\ref{3.7b})
it follows that any system
operator which commutes with $\hat Q_\rho$ and $\hat
Q_\rho^\dagger$ has to commute with the system constraint operators
$\hat\sigma^a_s$, and is thus
gauge invariant. Therefore the Green's functions obtained
from our description above
coincide with the system Green's functions of gauge invariant observables.

Finally we remark that the evaluation of
the Green's functions in a gauge theory
using the supersymmetric
quantization is not an easy task.
Though the problem of the supersymmetric boundary conditions
has been avoided, which is a great simplification from
the calculational point of view, there is still
a lack of a well elaborated techniques in quantum field theories to treat,
even in perturbation theory,
local
measures in the functional integral (see (\ref{3.9}))
and singular quantum potentials (see (\ref{3.10})) that usually
occur in unitary gauges.
Nonetheless it is possible to calculate the renormalized
Green's function using a modified loop expansion \cite{ts}.

\section{  The relation to gauge fixing.}
\setcounter{equation}0

Here we explain how the functional
integral (\ref{3.16}) is related to the functional integral
of a gauge fixed
Yang-Mills theory.  For simplicity we consider the model of section 3.
Let
$x=x_F$ be a generic configuration satisfying
the gauge condition $F(x)=0$.  We
assume that the gauge condition fixes all
continuous gauge arbitrariness, i.e.,
the stationary group of $x=x_F$ is trivial.
However, the equation $F(x)=0$ may
have many solutions related by discrete gauge transformations
(Gribov copies).
This gauge arbitrariness does not decrease the number of physical degrees of
freedom, but it does reduce the volume of the physical configuration space
spanned by $x_F$.  If all configurations $x_F$ form an Euclidean space
$I\!\!R^M$ and $S$ is a set of
residual discrete gauge transformations, then the
physical configuration space (a fundamental modular domain) is isomorphic to
$\Lambda_F\sim I\!\!R^M/S\subset  I\!\!R^M$.

Consider the change of variables $x_\rho=T_gx_F$, $z=\tilde T_g\chi\rho$,
$\eta=\tilde T_g\eta_\rho$, where the first relation determines the element
$T_g$ for a given $x_F$ and $x_\rho$, and $\tilde T_g$ is this element in the
fundamental representation.
The latter two relations define the new variables $z$ and $\eta$.
Introducing this change of variables in the
Hamiltonian (\ref{3.7}),
the ghost part assumes the form (\ref{ghh}), while in
the system Hamiltonian the Laplace operator $\hat p^2$ must be written in the
curvilinear coordinates $x_\rho=T_g x_F$.  If there were no ghosts in the
theory, the Laplace-Beltrami operator $\hat p^2$ would contain operators
$\hat\sigma^a_s$, corresponding to momenta conjugate to the group parameters
$\theta^a$.  Since the above change of variables involves the ghost variables,
all operators $\hat\sigma_s^a$ get replaced by $\hat\sigma^a_{gh}$ just as the
boson ghost constraints were replaced by $\hat\sigma^a_s+\hat\sigma^a_f$ in
(\ref{3.7}).  The functional integral for this Hamiltonian takes the form
(\ref{3.11}) where the measure $\rho^3$ is replaced by $\mu(x_F)$
($\int_0^\infty d\kappa_\rho\rho^3\int dx_\rho=\int d\kappa \int_{\Lambda_F}
dx_F\mu(x_F)$).  The measure $\mu(x_F)$ is the Faddeev-Popov determinant in the
gauge chosen.  The kernel (\ref{3.12}) contains a sum over all elements $T_s$
satisfying $F(T_sx_F)=0$.  Imposing supersymmetric boundary conditions for the
ghosts by choosing $\langle\langle\kappa,x_F|\psi\rangle\rangle_{ph}=
\delta_{gh}(\kappa)\psi^{ph}_s(x_F)$, the ghost integral is Gaussian and can be
done explicitly.  Boson and fermion determinants cancel and the prefactor is 1
due to the supersymmetric boundary conditions (see (\ref{bc})).  We therefore
end up with the functional integral for the system action in the gauge
$F(x)=0$.

For the Yang-Mills theory we can choose $F(A)=\partial_i A_i=0$ (Coulomb
gauge).  The above procedure yields the functional integral obtained in
\cite{sha2}.

\section{  Conclusions.}
\setcounter{equation}0

 The normal procedure of Lorentz covariant gauge fixing
has the disadvantage that it leads to Gribov copying.
The description of these
copies and the fundamental modular domain is extremely difficult and in the
continuum theory not well founded since it depends strongly on the functional
space chosen for the gauge potentials (the normal choice of $L^2$ forms a set
of zero measure in the functional integral) {\cite{sol}}.  However, to
calculate the non-perturbative Green's functions in covariant gauges one needs
all copies for a generic configuration satisfying the gauge condition
\cite{sha3}.  Given the complexity of these copies {\cite{sol}}, the
non-perturbative evaluation of Green's functions in covariant gauges seems to
be a hopeless task.

In the procedure developed above we have solved these difficulties.  Our
effective theory is Lorentz invariant and at the same time the Gribov problem
has been avoided.  The price we pay is in the appearance of an additional
scalar ghost field.  Furthermore we have encountered a difficulty with the
supersymmetric boundary
conditions to be imposed on the functional integral.
This difficulty can, however, be avoided in the path
integral for the Green's functions as we have shown.

We would like to stress that in the asymptotic free domain where perturbation
theory is valid, the normal approach of Lorentz covariant gauge fixing
{\cite{fad}} is more appropriate.  However, in the infra-red limit where the
non-perturbative aspects of the theory begin to dominate, the present formalism
should come into its own right.  One of its possible applications would
therefore be to study the infra-red behavior of the theory.  As the
present approach is gauge independent it makes it possible to extract the
infra-red properties of the theory due to the true dynamics of the
system (self-interactions of the gluons) and free from any influences coming
from the presence of the gauge dependent Gribov horizon in a gauge fixed
approach.  This would hopefully provide insight into the physical meaning of
the (non-perturbative) dynamically generated mass scale found in
{\cite{grib,zwan,stin}}.

\subsection*{ Acknowledgments.}

 We acknowledge support from the Alexander von Humboldt foundation and
the Foundation of Research development of South Africa.  We would also like to
thank H Kleinert for his
hospitality during the completion of a part of this work.

\section*{Appendix.}
\renewcommand{\theequation}{{\rm A}.\arabic{equation}}
\setcounter{equation}0

There exists an $N=2$ supersymmetric extension of the system dynamics.
To obtain it, one should find a
ghost extension $\hat P=\hat P^\dagger$ of the momentum $\hat p$ such that
\begin{equation}
[\hat P,\hat R]= [\hat P,\hat R^\dagger]=
[\hat P,\hat Q]= [\hat P,\hat Q^\dagger]=0
\end{equation}
and replace the $\hat p$ by it in all system operators.
One can convince oneself  that the operator
\begin{equation}
\hat P= \hat P^\dagger= \hat p-i
\beta(\hat x)( \hat c_2\hat c_1-\hat c_1^\dagger\hat c_2^\dagger+ \hat
b_2\hat b_1-\hat b_1^\dagger\hat b_2^\dagger)\,,
\end{equation}
with $\beta=-\omega^\prime/(2\omega)$, is the desired $N=2$ supersymmetric
extension of the momentum
operator.  An advantage of this extension is that we know the structure of
the eigenstates of the extended
Hamiltonian
\begin{equation}
|E\rangle\rangle=|N\rangle_{gh}^x\cdot|E^N\rangle_s\ ,
\end{equation}
where $|N\rangle_{gh}^x$ are the eigenstates of the
ghost number operator $N_{gh}$ and
\begin{equation}
(\hat H_s+\omega(\hat x)N)| E^N_s\rangle=E_s^N|E^N\rangle_s\ .
\end{equation}

The $N=2$ supersymmetry is not spontaneously broken in
this case as the groundstate
$|gs\rangle\rangle=|0\rangle^x_{gh}\cdot|gs\rangle_s $,
with $|gs\rangle_s$ the system groundstate, is
manifestly $N=2$ supersymmetric.

A disadvantage of this extension is the non-locality of the extended system
Hamiltonian that arises in the
field theory case.  Recall that $\omega^2(x)$ is
replaced by $-\tilde D^2_i(A) +m^2$ and, hence, $\beta$ would
involve the non-local operator $(-\tilde D^{2}_i(A)+m^2)^{-1}$.
For this reason we did not use this extension to
resolve the problem with the supersymmetric boundary conditions.


\begin{thebibliography}{1}

\bibitem{dir}
P. A. M. Dirac, {\it Lectures on Quantum Mechanics}
(Yeshiva University, N.Y., 1964).

\bibitem{ncom}
N. H. Christ and T. D. Lee, Phys. Rev. D {\bf 22}, 939 (1980),

L. V. Prokhorov, Sov. J. Part. Nucl. Phys. {\bf 13},1094(1982),

A. Ashtekar and G. T. Horowitz, Phys. Rev. D {\bf 26}, 3342 (1982).

\bibitem{fad}
L. D. Faddeev and V. N. Popov, Phys. Lett. B {\bf 25}, 30  (1967).

L. D. Faddeev, Theor. Math. Phys. {\bf 1},1  (1969).

\bibitem{sha1}
S. V. Shabanov, J. Phys. A {\bf 24}, 1199 (1991).

\bibitem{sha2}
S. V. Shabanov, Phys. Lett. B {\bf 255}, 398  (1991);
Phys. Lett. B {\bf 318}, 323 (1993).

\bibitem{bv}
I. A. Batalin and G. A. Vilkovisky, Phys. Lett. B {\bf 69}, 309 (1977).

\bibitem{grib}
V. N.  Gribov, Nucl. Phys. B {\bf 139}, 1 (1978).


\bibitem{sing}
I. M. Singer, Commun. Math. Phys. {\bf 60}, 7 (1978).

M. F. Aiyah and T. D. S. Jones, Commun. Math. Phys. {\bf 61}, 97 (1978).

\bibitem{sol}
M. A. Soloviev, Theor. Math. Phys. (USSR) {\bf 78}, 117 (1989).

\bibitem{zwan}
D. Zwanziger, Nucl. Phys. B {\bf 345}, 461 (1990).

\bibitem{ber}
C. Bernard, C. Parrinello and A. Soni, Phys. Rev. D {\bf 49}, 1585 (1994).

\bibitem{sch}
F. G. Scholtz and G. B. Tupper, Phys. Rev. D {\bf 48},  1792 (1993).

\bibitem{prok1}
L. V. Prokhorov, Sov. J. Nucl. Phys. {\bf 39}, 496 (1984).

\bibitem{sha3}
S. V. Shabanov, Mod. Phys. Lett. A {\bf 6}, 909 (1991).

\bibitem{ts}
G. B. Tupper and F. G. Scholtz, HEP preprint HEP-TH/9501089 (1995).

\bibitem{stin}
M. Stingl, Phys. Rev. D {\bf 34}, 3863 (1986).


\end{thebibliography}
\end{document}